\documentclass[
    aps,
    prx,
    letterpaper,
    nobalancelastpage,
    twocolumn,
    superscriptaddress,
    nofootinbib,
    longbibliography
]{revtex4-2}

\usepackage{graphicx}
\usepackage{amsmath}
\usepackage{amssymb}
\usepackage[english]{babel}
\usepackage{color}
\usepackage[version=4]{mhchem}
\usepackage[hidelinks]{hyperref}
\usepackage{dsfont}
\usepackage{multirow}
\usepackage{booktabs}
\usepackage{array}

\usepackage{soul}
\newcommand{\ket}[1]{{| {#1} \rangle}}
\newcommand{\bra}[1]{{\langle {#1} |}}

\newcommand{\UIUC}{
    Department of Physics,
    The University of Illinois at Urbana-Champaign,
    Urbana, IL 61801, USA
}

\renewcommand{\cite}[1]{\mbox{\citep{#1}}}

\addto\captionsenglish{}
\addto\captionsenglish{}

\DeclareUnicodeCharacter{2212}{-}

\begin{document}

\title{Active cancellation of servo-induced noise on stabilized lasers via feedforward}
\author{Lintao Li}\email{ltli@illinois.edu}
\affiliation{\UIUC}
\author{William Huie}
\affiliation{\UIUC}
\author{Neville Chen}
\affiliation{\UIUC}
\author{Brian DeMarco}
\affiliation{\UIUC}
\author{Jacob P. Covey}\email{jcovey@illinois.edu}
\affiliation{\UIUC}

\begin{abstract}
    Many precision laser applications require active frequency stabilization. However, such stabilization loops operate by pushing noise to frequencies outside their bandwidth, leading to large ``servo bumps" that can have deleterious effects for certain applications. The prevailing approach to filtering this noise is to pass the laser through a high finesse optical cavity, which places constraints on the system design. Here, we propose and demonstrate a different approach where a frequency error signal is derived from a beat note between the laser and the light that passes through the reference cavity. The phase noise derived from this beat note is fed forward to an electro-optic modulator after the laser, carefully accounting for relative delay, for real-time frequency correction. With a Hz-linewidth laser, we show $\gtrsim20$ dB noise suppression at the peak of the servo bump ($\approx250$ kHz), and a noise suppression bandwidth of $\approx5$ MHz -- well beyond the servo bump. By simulating the Rabi dynamics of a two-level atom with our measured data, we demonstrate substantial improvements to the pulse fidelity over a wide range of Rabi frequencies. Our approach offers a simple and versatile method for obtaining a clean spectrum of a narrow linewidth laser, as required in many emerging applications of cold atoms, and is readily compatible with commercial systems that may even include wavelength conversion.       
\end{abstract}
\maketitle

\section{Introduction}\label{sec: Intro}
The use of cold atom systems in burgeoning quantum science applications is pushing the limits of laser engineering. Many applications now require long-term frequency stability (sub-Hz) but are also sensitive to the fast frequency noise ($\sim$MHz). Nowhere is this more apparent than in the use of long-lived optical metastable states, which, in addition to their rich history in optical atomic clocks~\cite{Ludlow2015}, are gaining prominence in quantum computing~\cite{Monz2016,Madjarov2020,Schine2021,Yang2021,Allcock2021,Erhard2021,Chen2022,Wu2022,Okuno2022,Pagano2022,Ringbauer2022}, quantum simulation~\cite{Kokail2018,Joshi2020,Choi2021}, and quantum networking~\cite{Casabone2013,Lee2019,Covey2019b,Covey2019c,Huie2021}. Additionally, programmable entanglement in atomic clocks~\cite{Nichol2021,Schine2021,Marciniak2022} for quantum-enhanced precision~\cite{Gil2014,Kessler2014,Komar2014,Pezze2018,Kaubruegger2019} merges the requirements of optical metrology with quantum computing and networking. Atom-laser coherence at the second scale is required to fully leverage such metastable states, yet short-term stability is crucial for, e.g., gate operations and short term stability of optical atomic clocks limited by Dick noise~\cite{Oelker2019}. Moreover, the recent progress with the use of Rydberg states~\cite{Saffman2010} for programmable entanglement of neutral atoms~\cite{Levine2018,Levine2019,Graham2019,Madjarov2020,Schine2021,Ma2022} enables protocols with similar requirements, albeit with shorter timescales.

\begin{figure}[t!]
    \centering
    \includegraphics[width=0.5\textwidth]{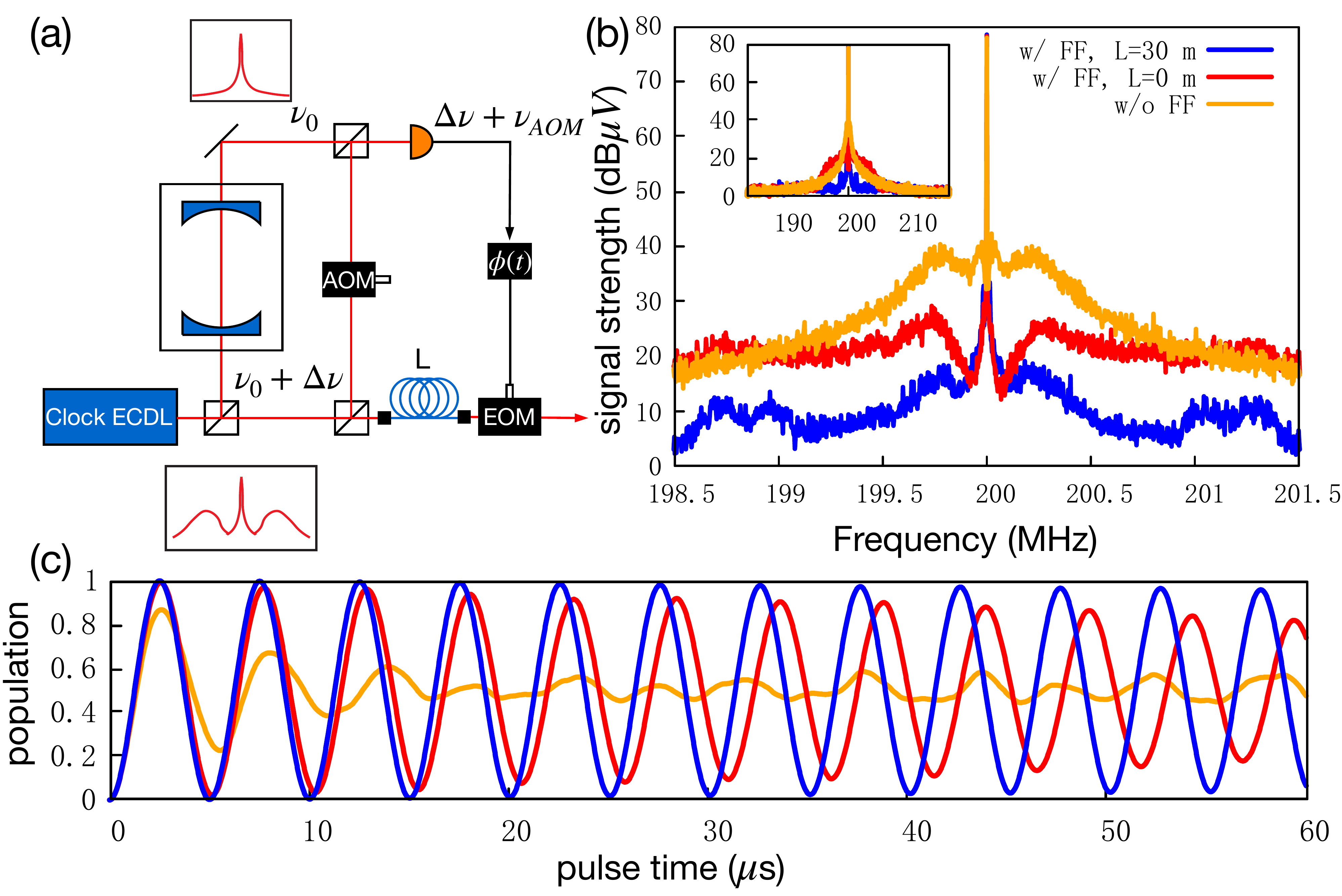}
    \caption{
        \textbf{Overview of the scheme and main results.} (a) An External Cavity Diode Laser (ECDL) is locked to a high finesse optical cavity, which produces ``servo bumps" on its spectrum. The transmitted light through the cavity serves as a spectral filter to remove these bumps, and the real-time frequency/phase deviation of the laser with respect to the cavity-transmitted reference is obtained via a beat note between the two. This signal is then fed forward to an electro-optic modulator (EOM), possibly with an optical fiber delay line added onto the light, to cancel the frequency deviations. (b) Laser spectra with (red and blue) and without (yellow) the feedforward (FF) applied to the laser. The blue (red) data shows the case where a 30 m fiber delay line is used (not used) prior to the EOM. The inset shows the same data over a wider frequency range. The resolution bandwidth is 100 Hz for all the measurements in this plot. (c) Simulations of driven two-level atoms under the three laser noise cases from (b) using a Rabi frequency of $\Omega=2\pi\times200$ kHz, which is near the peak of the servo bump.
        \label{Overview}
    }
\end{figure}

In recent years, it has become clear that laser noise at the qubit drive frequency is particularly deleterious~\cite{Levine2018,deLeseleuc2018}. For stabilized lasers, broad noise peaks from $\approx\pm0.1$ to 1 MHz, referred to as ``servo bumps", are inevitable due to the finite loop bandwidth. Therefore, the need for stabilized lasers with spectral filters to remove this high-frequency noise has become ubiquitous, with the pervasive approach based on using an optical cavity as a spectral filter~\cite{Martin2013,Levine2018} -- typically the same cavity to which the laser is stabilized. However, additional gain stages such as injection-locked diode lasers and tapered amplifiers are required when using the transmitted light through the cavity since the transmitted power $P_0\approx10$ $\mu$W is limited in practice by the power build-up $\mathcal{F}\times P_0$, where $\mathcal{F}$ is the cavity finesse. This limitation is particularly problematic in the ``ultrahigh" finesse ($\mathcal{F}\gtrsim100\,000$) range that is required for ``clock" lasers with fractional stability of $\lesssim10^{-15}$ at one second~\cite{Ludlow2015}, where the transmitted power may not even be high enough to injection-lock a diode laser. Moreover, the cavity transmission technique is challenging for laser systems that involve frequency conversion -- particularly commercial systems -- such as second-harmonic generation, which is common for neutral ytterbium, mercury, and cadmium; ionic aluminum, beryllium, and magnesium; and Rydberg transitions of essentially every species.

Here we demonstrate an alternative approach to spectral filtering of servo bumps based on active noise cancellation via a feedforward technique. By using the cavity transmitted light to generate a beat note with the laser~\cite{Ivanov2006,Schmid2019}, the frequency deviation of the laser from the cavity-filtered reference can be corrected in real time with an active optical device such as an electro-optic modulator (EOM)~\cite{Hall1984,Aflatouni2012} [see Fig.~\ref{Overview}(a)]. With optimal signal delay compensation using an optical fiber, we demonstrate that this technique can suppress noise up to $f_\text{nc}\approx5$ MHz by at least 3 dB below that of the original laser, and that the peak of the servo bump ($f_\text{p}\approx250$ kHz) is suppressed by $\approx$22 dB. Noise above $f_\text{nc}$ is not adversely affected and remains at the level of the original laser, and $f_\text{nc}$ is limited only by the electronic circuitry. Without including an optical fiber, $f_\text{nc}\approx1$ MHz and the servo bump peak is suppressed by $\approx15$ dB. [See Fig.~\ref{Overview}(b).] Moreover, we demonstrate that this technique does not come at the expense of long-term stability, even in the case with a fiber delay line, by showing a Fourier-limited width of $\lesssim20$ mHz for the coherent peak of the beat note. Finally, we simulate the dynamics of a two-level system under a time-dependent drive based on our data, showing an enormous improvement [Fig.~\ref{Overview}(c) shows Rabi frequency $\Omega\approx f_\text{p}$]. Our approach is simple and versatile, provides noise suppression out to high enough bandwidths to be relevant for Rydberg excitation, and can readily be applied to complex and commercial laser systems that involve both high-finesse cavities and frequency conversion.

\section{Overview of the technique}\label{Technique}
We now provide a more detailed description of the technique and the circuitry, as shown in Fig.~\ref{Figure2}. Further details can be found in Appendix~\ref{Circuit}. We send the light from the ECDL through an acousto-optic modulator (AOM) and interfere half of it with half of the light collected after the cavity on a 50:50 beamsplitter (BS), generating a beat note at 200 MHz that is measured with a photodiode (PD1 in Fig.~\ref{Figure2}). Note that the amplitude of this beat note scales as the product of the electric field of both sources, so even a small cavity-transmitted power (we use $\approx4$ $\mu$W for beating with a $\approx40$ $\mu$W laser from the ECDL in the experiment) provides sufficient signal. The cavity-transmitted power can be further reduced by scaling up the ECDL laser power by the same factor.

\begin{figure}[t!]
	\centering
	\includegraphics[width=0.5\textwidth]{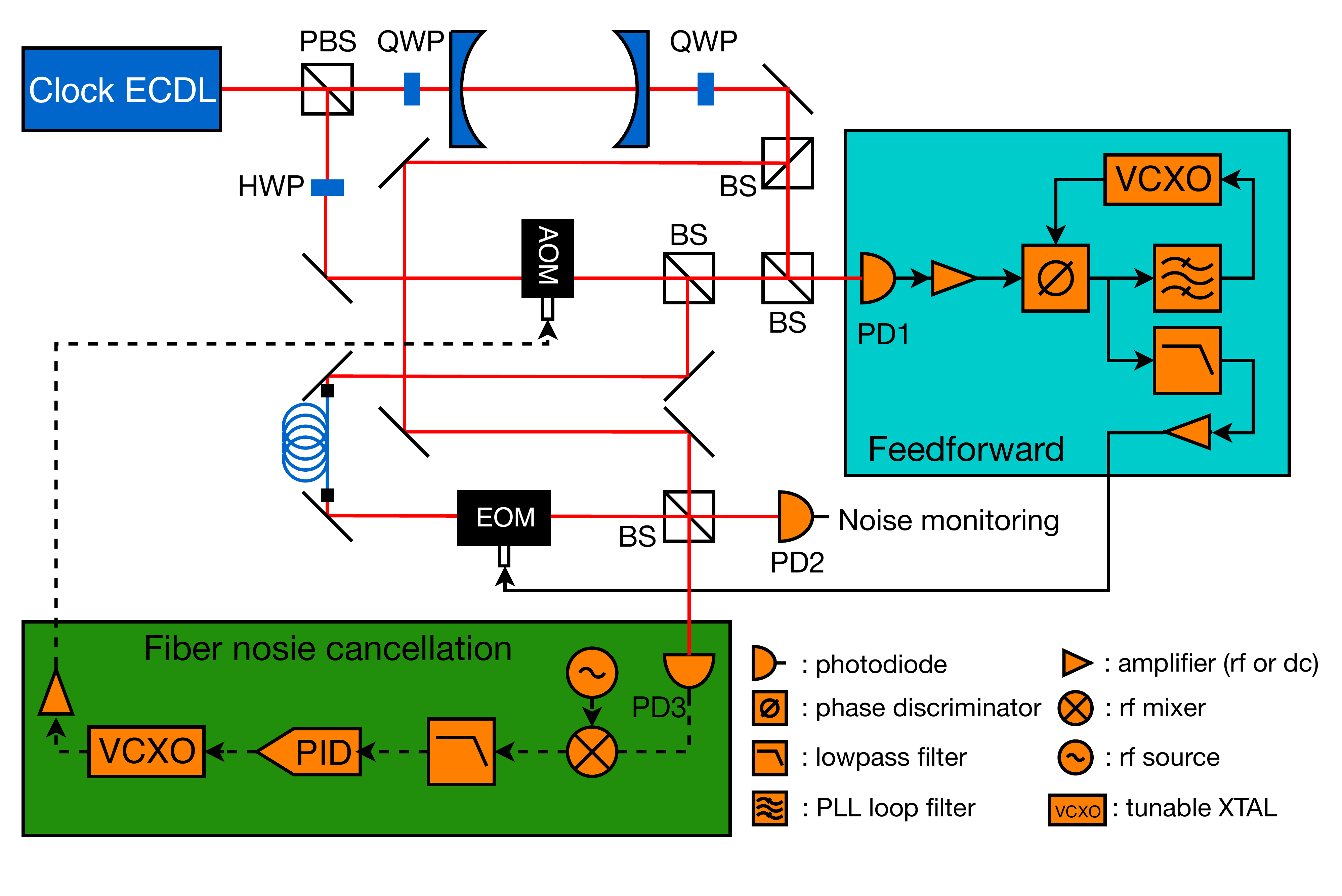}
    \caption{
        \textbf{Detailed schematic}. The ECDL light and the cavity-transmitted light are used in two places for our measurements, so each is split with a 50:50 beamsplitter (BS). An acousto-optic modulator (AOM) is used to create a 200-MHz beat note between the ECDL and the cavity-transmitted light on photodiode (PD) 1 by interfering them on a BS. This signal is used to cancel frequency deviations from the cavity-transmitted reference through the use of the feedforward phase correction circuit shown in the teal box. After the AOM, the laser light is sent through an optical fiber delay line before going to an electro-optic modulator (EOM). The phase signal from the circuit in the teal box is applied to the EOM. Then, this light is interfered with the cavity-transmitted light on a BS and monitored on two PDs. PD2 is used for measurements, and represents what would be sent to the atoms. PD3 is used to measure slow fiber-induced frequency drifts which are then corrected with the fiber noise cancellation circuit (green box). This feedback system corrects the frequency applied to the AOM to compensate these slow drifts. The feedback and feedforward work together to maintain the highly-coherent peak of the cavity-stabilized laser while simultaneously removing fast noise from the servo bumps.  
        \label{Figure2}
    }
\end{figure}

The other half of each source is used to perform the feedforward for noise cancellation and to measure the result. The laser light after the AOM is sent through an optical fiber delay line and then into an EOM. We use a free-space EOM rather than a waveguide EOM both because it can handle substantially higher optical power and because it has lower residual amplitude modulation (see below). The output is interfered with the cavity transmitted light on a BS and both ports are monitored with a PD (PD2 and PD3 in Fig.~\ref{Figure2}). The signal from PD1 is converted to a time-dependent phase deviation, which is amplified and sent to the EOM with proper polarity. The phase locked loop (PLL)-based phase detection circuit has a working frequency range from around 200 Hz to 4.8 MHz (see Appendix~\ref{Circuit}). The wide band amplifier which drives the EOM has a bandwidth exceeding 15 MHz. The circuitry, from PD1 to the EOM, introduces a delay of $\tau_\text{e}\approx140$ ns (see Appendix~\ref{Circuit}), so we add an optical delay $\tau_\text{o}$ using an optical fiber. As discussed below, we identify $L=30$ m as the optimal fiber length such that $\tau_\text{e}\approx\tau_\text{o}$.

The addition of the fiber delay helps to optimally cancel high frequency noise, but it is well known that thermal and acoustic fluctuations introduce low frequency noise to optical signals passing through fibers, which can be canceled with active feedback~\cite{Ma1994}. We use PD3 to monitor this slow noise by mixing the 200-MHz signal with a radiofrequency (rf) source of the same frequency, thereby creating a DC signal that is used in a PID servo loop with the frequency-tunable rf source that drives the AOM. Therefore, the frequency of the laser is actively adjusted to compensate for the slow, fiber-induced drifts. Due to this fiber noise cancellation, the rf frequency of the AOM varies slowly with time. Accordingly, the beat signal on PD1 will not only have the high frequency servo bumps from the ECDL but also the slowly-varying phase noise from the AOM. Since the feedforward circuit will only response to phase noise higher than 200 Hz (see Appendix~\ref{Circuit} for detailed description), the slow fiber noise cancellation does not adversely affect the function of the feedforward circuit. In this way, we can preserve the long-term stability of the laser obtained by locking it (feedback) to the high finesse cavity while simultaneously removing high frequency, servo-induced noise via feedforward.

Our specific implementation of the fiber noise cancellation can only be applied when the fiber is spooled such that the output is proximal to the input and other optical paths. In the case where a fiber is used to span some distance, more standard implementations of fiber noise cancellation can be employed using the reflection from the fiber output tip to generate an inteferometer~\cite{Ma1994}. 

Finally, we emphasize that this feedforward technique, by construction, cannot filter noise better than the cavity does -- although it comes close over a wide bandwidth that covers most of the laser phase noise. Specifically, the cavity used in this work has a modest finesse (see Appendix~\ref{LaserSystem}), so the results discussed below for the feedforward data with the $L=30$ m delay length probably exceed the quality of the cavity-transmitted spectrum. That is, we assume the cavity-transmitted signal has no noise. However, our approach is intended for ultrahigh finesse cavity systems for which the cavity-transmitted noise becomes negligible.

\section{Results}\label{Results}
\begin{figure*}[t!]
	\centering
	\includegraphics[width=0.75\textwidth]{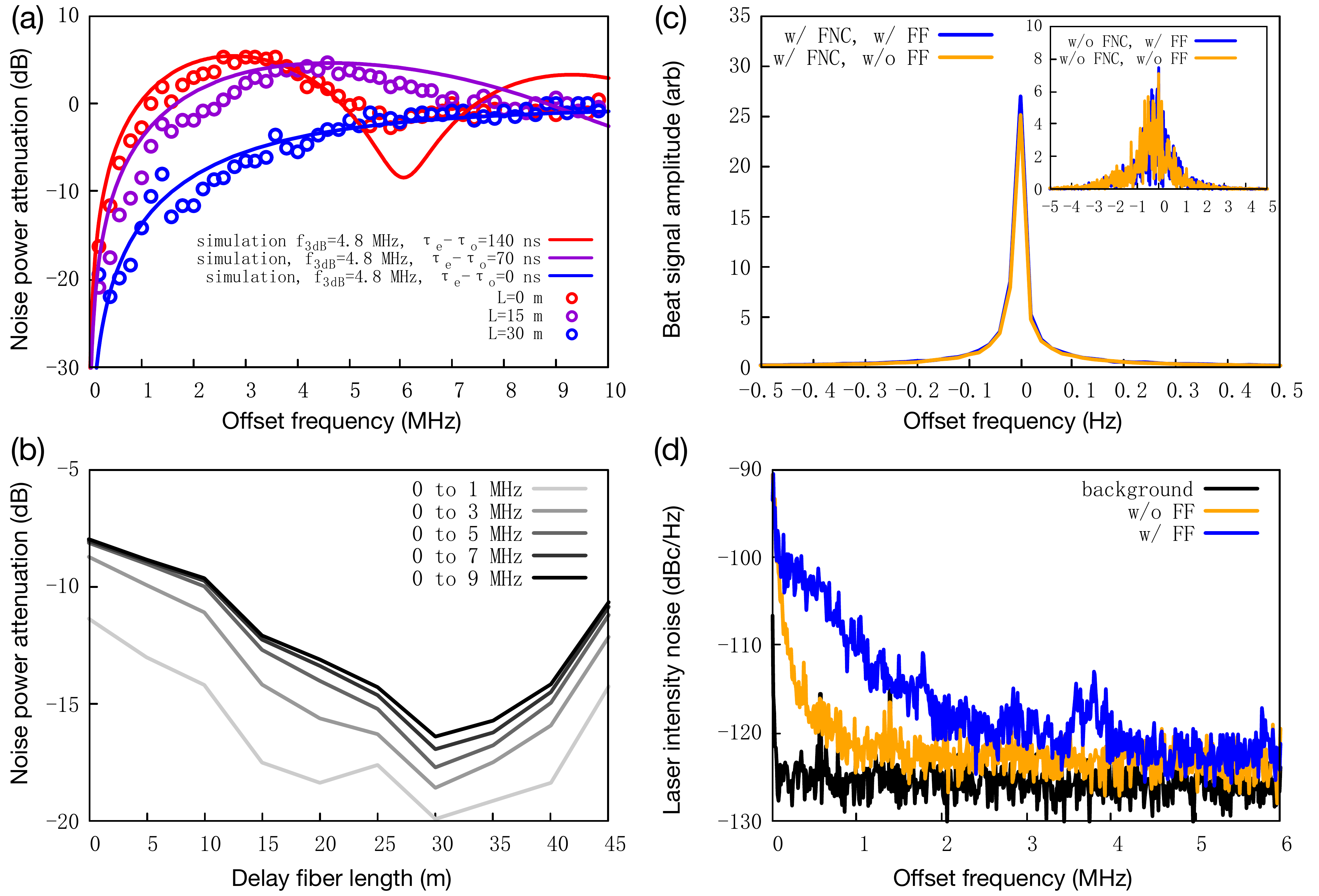}
    \caption{
        \textbf{Results}. (a) Noise difference spectra $\mathcal{I}(f)$ between the case without feedforward enaged and cases with it engaged versus frequency for several optical fiber delay length $L$ (red: 0 m, purple: 15 m and blue: 30 m). Negative values show reduced noise and positive values show increased noise. The points are measured data while the lines are simulations based on the circuit transfer function (see Appendix~\ref{Circuit}). (b) The integrated noise difference $\mathcal{I}(f)$ from (a) between $f=0$ and $f'$ for several values of $f'\in\{1,...,9\}$ MHz versus optical fiber delay length $L$, showing a clear minimum at $L=30$ m. (c) The noise spectrum over a narrow range with $L=30$ m, with and without feedforward engaged. The main figure shows the case with a slow feedback loop to the AOM to cancel fiber noise, giving a Fourier-limited coherent peak. This indicates that our technique maintains the intrinsic laser linewidth. The inset shows the case without canceling fiber noise. (d) The laser intensity noise with and without applying the feedforward, and the background. We see that the feedforward makes the intensity noise higher due to residual amplitude modulation from the EOM. We show below that this effect is negligible. 
        \label{Figure3}
    }
\end{figure*}

To quantify our results, we compare at each frequency the spectral density of the feedforward signal $\mathcal{I_\text{w/}}(f)$ to the case without engaging the feedforward $\mathcal{I_\text{w/o}}(f)$. That is, $\mathcal{I}(f)=\mathcal{I}_\text{w/}(f)-\mathcal{I}_\text{w/o}(f)$. Hence, $\mathcal{I}<0$ in the spectral region where the feedforward reduces the noise, and $\mathcal{I}>0$ in regimes where the feedforward increases the noise. The bandwidth in which the feedforward reduces the noise depends on the relative delay between the optical and electrical signals. Figure~\ref{Figure3}(a) shows $\mathcal{I}(f)$ for the case of $L=0$, $15$, and $30$ m, corresponding to $\tau_\text{o}\approx0$, $70$, and $140$ ns, respectively. While all three cases show $\mathcal{I}<0$ below $\approx$1 MHz, the $L=0$ and $L=15$ m cases have $\mathcal{I}>0$ beyond this frequency before approaching zero. In the case of $L=0$, our numerical model based on the circuit transfer function (see Appendix~\ref{TransferFunction}) predicts that there will be a dip below zero near 6 MHz where the relative phase delay is roughly $2\pi$, such that the feedforward can successfully cancel noise. Our data does not corroborate this dip, which we believe is because there is limited noise to cancel at 6 MHz due to the profile of the servo bump being primarily below 1 MHz. For $L=30$ m our data is in agreement with our numerical modeling to suggest that the relative delay is small, offering optimal performance of the feedforward method. The region where $\mathcal{I}<0$ is maximized under this condition, and the signal asymmtotes to zero rather that first going positive.

To identify the optimal fiber delay length $L$, we define a figure of merit $\mathcal{T}(f')$ as the integrated noise under $\mathcal{I}(f)$ from $f=0$ up to a chosen bandwidth $f=f'$. Figure.~\ref{Figure3}(b) shows $\mathcal{T}(f')$ for several values of $f'\in\{1,...,9\}$ MHz, where there are minima at $L=30$ m for all values of $f'$. This is in good agreement with our understanding based on the transfer function of our system (see ~Appendix~\ref{TransferFunction}), suggesting that there is $\tau_\text{e}\approx140$ ns of electronic delay. For $L=30$ m, $\tau_\text{o}=nL/c\approx140$ ns for optical fiber with index of refraction $n=1.44$.   

As noted above, the fiber delay line can introduce noise due to acoustic and thermal path length fluctuations. Thus, the use of a fiber to compensate electronic delay introduces low frequency noise that must be removed with active feedback. Figure~\ref{Figure3}(c) shows beat note spectra similar to those in Fig.~\ref{Overview}(b) except over a much narrower frequency range. We consider the optimal case of $L=30$ m. Figure~\ref{Figure3}(c, inset) shows the situation without removing the fiber-induced noise for two cases: where the feedforward that removes the high frequency noise (servo bumps) is and is not engaged. There is no apparent difference on this scale. Figure~\ref{Figure3}(c, main figure) shows the same two cases, where now the fiber-induced noise is removed by feeding back to the AOM as described above. Here, the spectra are Fourier limited and there is no apparent broadening due to the fiber. We see that the peak corresponding to the case where the feedforward is engaged has a higher amplitude than the peak where feedforward is not engaged. This is due to the noise cancellation process that actively transfers the spectral weight of the laser noise back into the coherent peak. Indeed, the increased amplitude shown here is comparable with the spectral weight under the servo bumps and the laser power is unchanged during the measurement.

Note that this spectrum is not a true measure of the laser linewidth since the coherence length is much longer than the path length difference in the interferometer arms. We believe from other data that our laser has $\gtrsim1$ Hz linewidth (see Appendix~\ref{LaserSystem} for details about the laser system). 

We also consider the intensity noise of the laser with and without applying the feedforward. Figure~\ref{Figure3}(d) shows the intensity noise of the background as well as the laser light in both cases. We see that the case with the feedforward applied has increased noise within $\approx$4 MHz compared to the case without the feedforward applied. We attribute this to residual amplitude modulation (RAM) on the EOM, in which frequency modulation is partially mapped onto amplitude modulation, typically due to an imperfect input polarization. This problem is more prevalent with waveguide EOMs, and our findings underline our decision to use a free-space EOM. We believe that improvements to our setup would further reduce the measured RAM. To quantify the significance of this added intensity noise, we turn to numerical modeling.  

\begin{figure}[t!]
	\centering
	\includegraphics[width=0.4\textwidth]{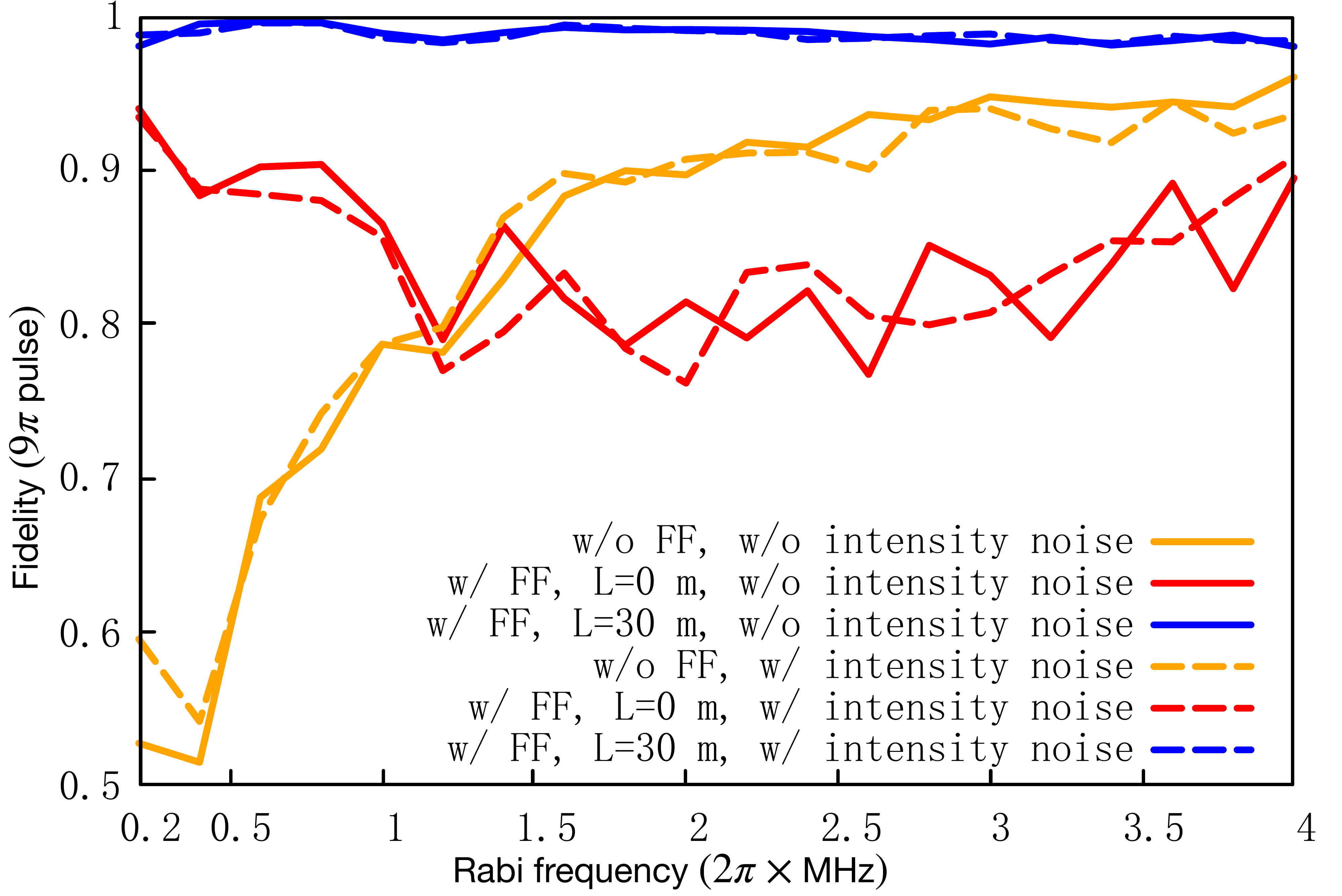}
    \caption{
        \textbf{Simulation of a driven two-level atom}. The fidelity after a $9\pi$-pulse (fifth maximum of the population transfer fraction) of Rabi dynamics with time-dependent drives based on measured noise spectra (see Appendix~\ref{TwoLevel}) is shown for the cases without feedforward (solid yellow) and with feedforward for $L=0$ and $L=30$ m (solid red and blue). We see that the $L=30$ m case (blue) has excellent fidelity over the entire bandwidth. The $L=0$ m case (red) is improved compared to the no-feedforward case below $\approx1$ MHz, but degrades substantially at higher frequency. The case without feedforward (solid yellow) returns to high fidelity for frequencies well above the servo bump regime. We also consider the case where intensity noise is included (dashed lines). We see that the intensity noise is negligible compared to the frequency noise. 
        \label{Figure4}
    }
\end{figure}

We simulate a driven two-level atomic system under the present of the measured noise to gauge the effect of our noise cancellation system on atomic dynamics (see Appendix~\ref{TwoLevel} for details). First, we consider only frequency noise. We specifically focus on the cases of the feedforward engaged with $L=0$ and $L=30$ m and the feedforward not engaged. Figure~\ref{Figure4} shows the $9\pi$ pulse fidelity, which is the fifth maximum population transfer fraction of the Rabi evolution, versus the Rabi frequency used in the simulation. We observe that Rabi frequencies in regions of high noise give particularly poor fidelity, and thus these results are qualitatively similar to the inverse of the noise spectra shown in Fig.~\ref{Overview}(b). [Figure~\ref{Overview}(c) shows time traces of these Rabi evolutions for a Rabi frequency of $\Omega=2\pi\times200$ kHz.] Correspondingly, we see in Fig.~\ref{Figure4} that the case without feedforward has poor fidelity between 0.2 MHz and $\approx1$ MHz, which then asymptotically approaches unity well beyond the servo bump bandwidth. The case of feedforward with $L=0$ m is significantly improved at low frequency, but this improvement is lost beyond $\approx1.3$ MHz where the relative delay time causes the feedforward to degrade the fidelity. The case of feedforward with $L=30$ m has excellent fidelity over the entire frequency range since noise is efficiently canceled within the servo bump bandwidth and noise is not added outside this bandwidth.

Figure~\ref{Figure4} also includes simulations with the measured intensity noise from Fig.~\ref{Figure3}(d). (See Appendix~\ref{TwoLevel} for details.) We consider both frequency noise and intensity noise for the two cases with feedforward applied. In practice, we believe the frequency noise may be synchronized with the intensity noise, but the two are uncorrelated in our simulations. We find that the intensity noise is irrelevant compared to the frequency noise over the entire range of $\Omega$ considered. This can be understood by considering the similarity between the amplitude and the phase modulation for small index. We can express a phase-modulated signal with the phase variation defined by $\Delta\phi\to0$ as $\textrm{sin}(\omega t+\Delta\phi)\approx\textrm{sin}(\omega t)+\Delta\phi\cdot\textrm{cos}(\omega t)$, where $\omega$ is the carrier frequency, and the second term is amplitude-modulated by the phase variation $\Delta\phi$. By converting the curves in Figure~\ref{Overview}(b) into phase noise (see Appendix~\ref{TwoLevel}), we find the phase noise is always higher than -90 dBc/Hz even for the best configuration (with feedforward and 30 m fiber delay). This phase noise level is still much higher than the intensity noise in Fig.~\ref{Figure3}(d).

\section{Concluding discussion}\label{Outlook}
We have presented a simple and versatile alternative to the cavity filtration-based approach to removing servo-induced noise on stabilized lasers. Our technique is based on measuring the real-time frequency deviation from the cavity resonance via a beat note between the cavity-transmitted light and the laser, which is then fed forward to an active device. We find that an EOM after a $L=30$ m optical fiber delay line to compensate the $\tau_\text{e}\approx140$ ns electronic delay provides noise suppression of $\approx22$ dB within the servo bump and can suppress noise with a bandwidth as high as $\approx5$ MHz. We also demonstrate that this technique can be applied without compromising the long-term stability of the laser by demonsrating a Fourier-limited, $\lesssim20$ mHz linewidth beat note for the coherent peak. This is accomplished by adding a fiber noise cancellation feedback loop, and our technique seamlessly combines the feedforward for high frequency noise removal with the feedback for low frequency noise removal.

In cases where the laser system includes frequency doubling or quadrupling from the wavelength referenced to the optical cavity, the only required change to our approach -- if it is to be applied at the final output -- is to apply a commensurate factor to the frequency modulation applied to the EOM downstream. More generally, any other transfer function can be added to the electrical circuit to match a transfer function in the laser system such as, for instance, sum or difference frequency generation that could be combined with second harmonic generation (SHG). Given the large bandwidths of single-pass nonlinear optics and resonant SHG cavities, we do not anticipate significant effects on the feedforward bandwidth even in such complex laser systems. Finally, for UV systems where it may be impractical to have an optical fiber, we point out that the fiber delay line could come before a frequency doubling stage, for instance. In particular, the fiber and EOM could be before the doubling stage(s) and even before an amplification stage with saturable gain, obviating the already negligible added intensity noise due to the EOM RAM. 

Additionally, we suggest a non-optical variant of our feedforward technique specifically for the case of driving ground-Rydberg transitions. Instead of using the real-time frequency deviation to apply a correction to the light via the EOM, one could instead apply the correction to the atomic transition frequency. In particular, a microwave field detuned from a nearby, opposite-parity Rydberg state could be used to Stark-shift the target Rydberg state~\cite{Kumar2022}. Amplitude modulation of this microwave field would thereby modulate the optical frequency of the ground-Rydberg transition. This technique could, in principle, have higher bandwidth than the EOM.

In conclusion, we believe that our fast feedforward technique offers new opportunities for state-of-the-art laser engineering, and may find application in intensity noise mitigation as well. Our approach provides increased access to ultranarrow lasers with a clean spectrum free of servo bumps, approaching the true delta-function limit without the complexity associated with the cavity filtration technique -- especially for ultrahigh finesse cavity systems. Therefore, we believe our techniques will find application in myriad new quantum science directions that rely on optically metastable states and Rydberg states. 

\section*{Acknowledgments}
We thank Brett Merriman for helpful discussions and Jun Ye for critical reading of the manuscript. We acknowledge funding from
the NSF QLCI for Hybrid Quantum Architectures and Networks (NSF award 2016136), the NSF PHY Division (NSF award 2112663), the NSF Quantum Interconnects Challenge for Transformational Advances in Quantum Systems (NSF award 2137642), and the ONR Young Investigator Program (ONR award N00014-22-1-2311).\\

\setcounter{section}{0}
\twocolumngrid

\renewcommand\appendixname{APPENDIX}
\appendix
\renewcommand\thesection{\Alph{section}}
\renewcommand\thesubsection{\arabic{subsection}}



\section{Drive circuitry and simulation}
\label{Circuit}

The drive circuitry is divided into two parts, the phase measurement circuit and the driver for the free-space EOM.

As we have described in the Main Text, the beat signal on PD1 includes both the high frequency phase noise from the servo bump and the slow varied phase noise from the AOM. The working principle of the phase measurement circuit is based on using a low bandwidth PLL that tracks the slowly-varying phase noise but ignores the high frequency phase noise. Since the voltage-controlled oscillator tracks the slowly-varying phase noise from the AOM, only the servo bump phase noise will appear after the phase detector.

\begin{figure}[t!]
	\centering
	\includegraphics[width=0.48\textwidth]{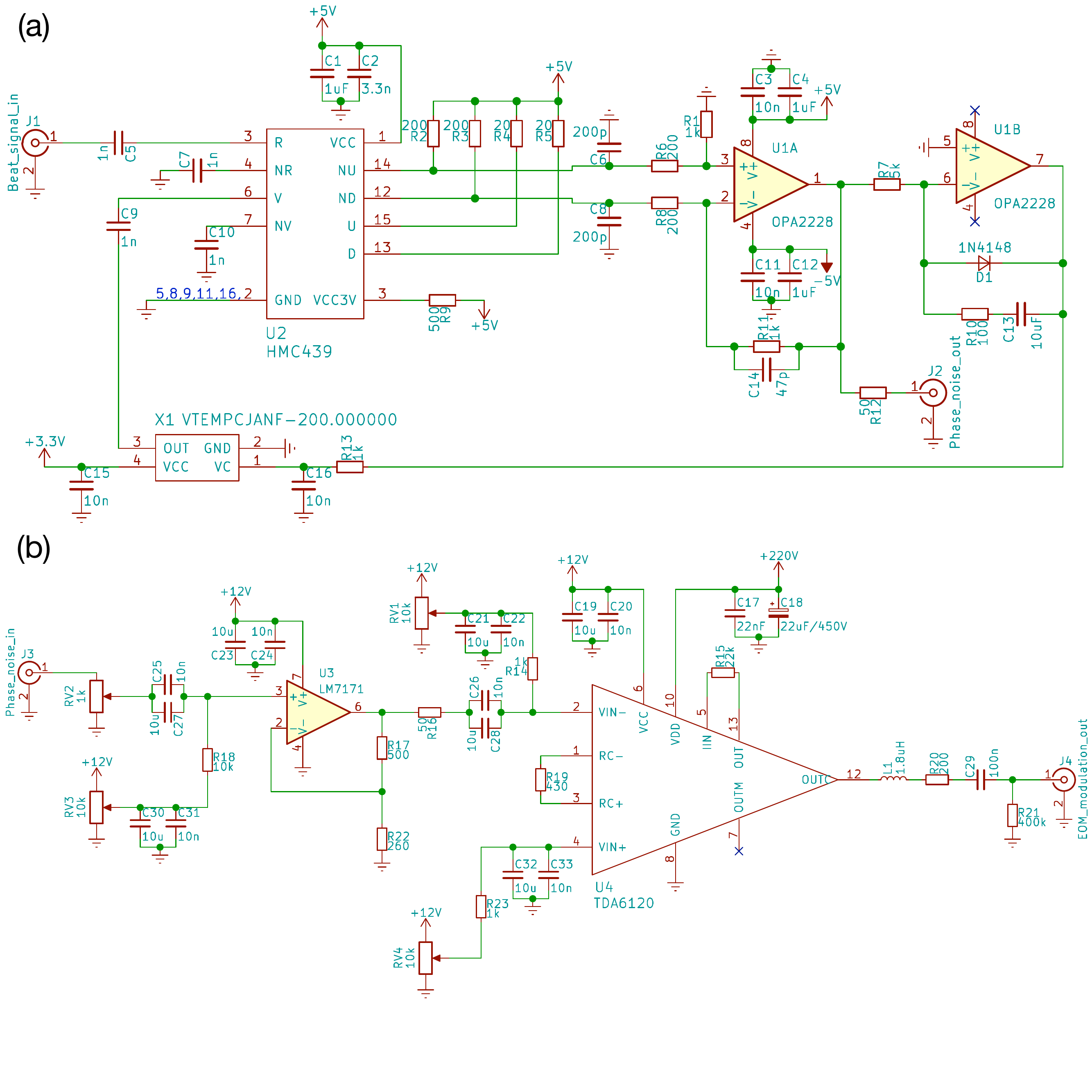}
	\caption{
        \textbf{Drive circuitry used in the feedforward noise canceling setup.}
        (a) The PLL-based phase detection circuit which converts the real-time phase deviation of the input beat signal into voltage. (b) The high bandwidth and high voltage wideband amplifier which amplifies the phase deviation signal to high voltage that matches the $V_\pi$ of the free-space EOM.
	\label{FigureSchematic}
    }
\end{figure}

The real-time phase measurement circuit is shown in Figure~\ref{FigureSchematic}(a). The phase compactor U2 compares the phase difference between the input beat signal J1 and the output of the voltage controlled crystal oscillator X1. The oscillator X1 has very low phase noise and serves as the reference oscillator for the phase comparison. Thus the output of U2 is mainly dependent on the phase noise of the input beat signal. Since the phase compactor only works for a phase difference ranging from $-\pi$ to $\pi$, the output of the phase compactor is fed back to X1 via a slow loop filter formed by U1B. Consequently, the slow fiber noise cancellation servo, the long term phase drifts, and the unwanted phase wraps will not affect the measurement of the phase noise. Finally, the wide band DC amplifier U1A converts the differential output of U2 to a single ended phase noise voltage signal which is sent to the EOM driver circuit.

The free-space EOM is preferred in this setup because of its ability to handle high power and good polarization stability. However, this type of EOM needs relatively high driving voltage compared with a waveguide EOM. A high speed and high voltage EOM driver is required to decrease the electronics delay and increase the modulation bandwidth. Figure~\ref{FigureSchematic}(b) shows the schematic of our homemade EOM driver that meets these requirements. The key component is a cathode-ray tube (CRT) driver (Philips Semiconductors, TDA6120Q) that was used to drive the cathodes of a CRT in High Definition TVs or monitors in past decades. The amplifier is tested up to a peak-to-peak voltage output of 160 V with a slew rate around $10\,000\textrm{ V}/\mu$s. A typical low-capacitance free-space EOM (Thorlabs, EO-AM-NR-C4) with smaller than 20 pF can be used since it is comparable to the capacitance of a CRT cathode. However, any long distance coax cable should be avoided due to the high parasitic capacitance.

The simulation of the phase calculation circuit is based on TINA-TI software~\cite{TI2008}, and the voltage-controlled crystal oscillator and the phase detector are replaced with an integrator which gives the same overall transfer function. The simulation indicates a lower frequency response of around 200 Hz and a high frequency response up to 4.8 MHz. For the step response, the simulation gives a bandwidth limited delay of around 130 ns; this delay together with the 10 ns propagation delay in the high voltage amplifier contributes the overall electronics delay $\tau_e\approx140$ ns.  The circuit simulation matches the measurement of the noise canceling factor calculated in Figure~\ref{Figure3}(a).

The physical concept underlying our phase-noise cancellation is the fact that all the noise that we want to remove comes from the feedback system, which is the intrinsic laser phase noise multiplied by the noise gain $[1+A(s)]^{-1}$ of the feedback loop, where $A(s)$ is the open-loop gain of the system. The function $[1+A(s)]^{-1}$ is also known as the noise shaping function which moves the laser phase noise at low frequency up to higher frequency~\cite{Pavan2017}. The information of the phase noise can either be derived from the error signal that modulates the laser frequency or by interfering the laser output with the cavity transmission~\cite{Schmid2019}.

\begin{figure}[t!]
	\centering
	\includegraphics[width=0.35\textwidth]{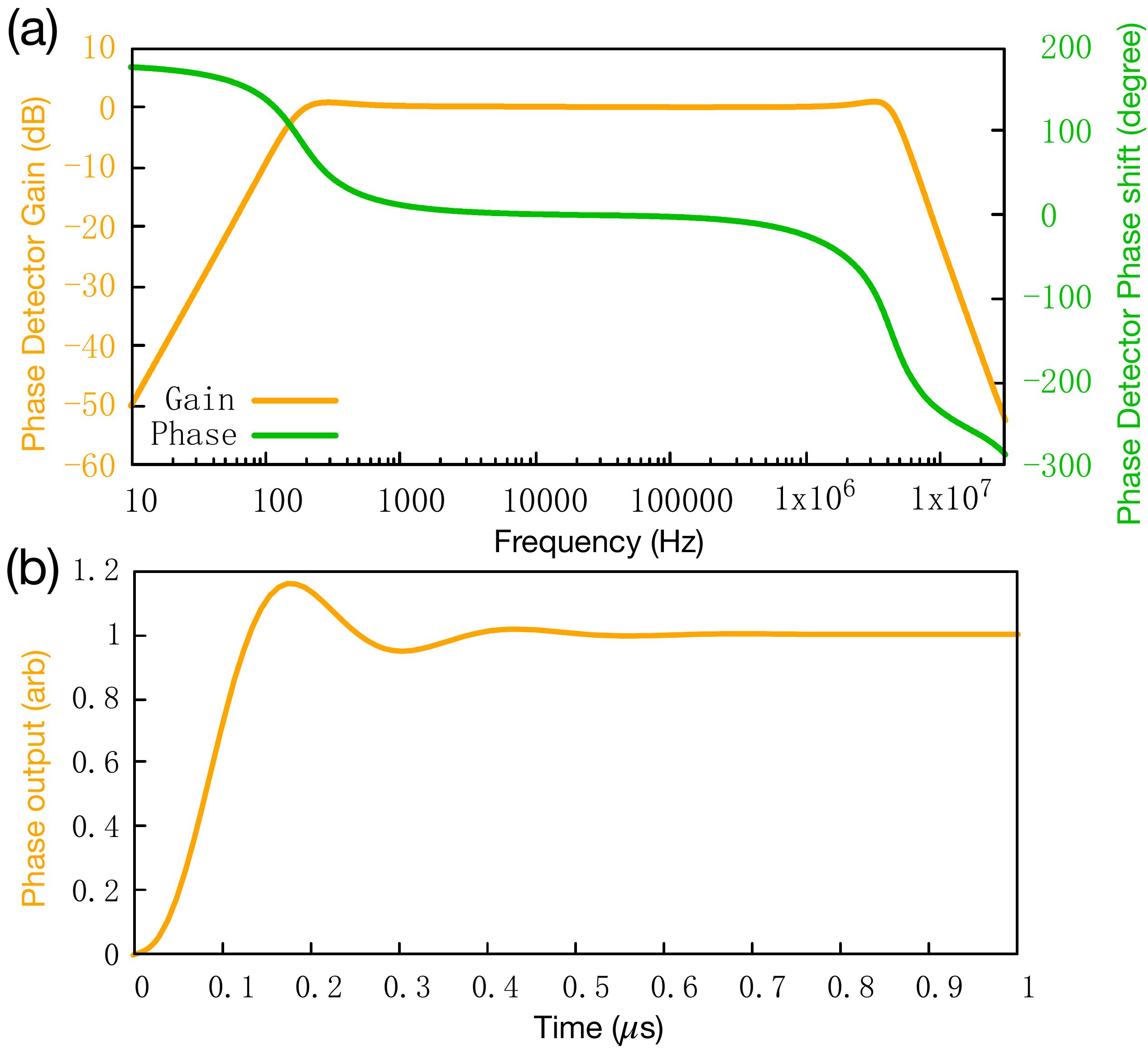}
	\caption{
        \textbf{Circuit simulation of the phase detection circuit.}
        (a) The Bode plot of the phase detection circuit's output versus the phase variation signal input. (b) The step response of the phase detection circuit's output.
	\label{FigureCircuitSimulation}
    }
\end{figure}

\section{Feedforward noise cancellation and the modeling}
\label{TransferFunction}

With the knowledge of the phase noise, we use the feedforward system following a high-finesse, cavity-locked laser to remove the servo bump. Compared with a feedback-based approach, a feedforward system gives us more flexibility and advantages for removing the phase noise. Firstly, in a feedforward system, the control variable is not error-based and thus there is no stability issue that is pervasive for a closed-loop system. Secondly, the delay $\tau_e$ of the phase noise measurement and frequency modulation device can be compensated by introducing the same amount of delay in the optical path. In a closed-loop system, the loop bandwidth of a feedback system is always limited by $1/4\tau_e$~\cite{Drever1983}.

Considering the finite bandwidth and the delay in the system, the transfer function of the feedforward phase noise canceling system can be written as~\cite{Oppenheim1996}
\begin{equation}
    H(s)=1-\frac{\boldmath{e}^{-(\tau_e-\tau_o)s}}{1+\frac{s}{2\pi f_{3dB}}}
,\end{equation}
where $f_{3dB}$ is the bandwidth of the feedforward system. The factor $\boldmath{e}^{-(\tau_e-\tau_o)s}$ describes the phase delay introduced by the fixed optical and electronics time delay. When $\tau_o = \tau_e$, the transfer function $H(s)$ gives a noise attenuation factor $f^2_\text{3dB}/(f^2_\text{3dB}+f^2)$, which is 0.5 when $f=f_\text{3dB}$ and gradually approaches 0 for higher frequency. For a system with infinite bandwidth, the feedforward can be understood as a self-heterodyne measurement with the fiber delay $\tau_o$ replaced by the overall delay of $\tau_e-\tau_o$~\cite{Okoshi1980, Richter1986}. The frequency response of the transfer function $H(s)$ under different $\tau_e-\tau_o$ is shown in Figure~\ref{Figure3}(a) and the dip in the red curve can be understood as the laser phase noise around 6 MHz being canceled by the measurement of the same phase noise around one period ($\approx140$ ns) before.

Finally, we note that a similar feedforward method is also employed in phase noise removal of the microwave band oscillator. A sub-sampling phase detector (SSPD) is used for measuring the phase noise and feed it forward to a variable delay line for canceling the phase noise~\cite{Nagam2018}.

\section{Modeling a driven two-level atom}
\label{TwoLevel}

In order to quantitatively evaluate the influence of the laser phase noise on the atomic state dynamics, we numerically solve the Schr\"odinger equation of a two-level system driven by the time-dependent phase factor $\phi(t)$ and coupling fluctuation factor $\epsilon(t)$. For zero laser detuning, the Hamitonian of the system is given by
\begin{equation}
    \hat{H}(t)=\frac{[1+\epsilon(t)]\Omega}{2} e^{-i\phi(t)}\ket{e}\bra{g}+\textrm{H.c.}
,\end{equation}
where $\Omega$ is the Rabi frequency, $\phi(t)$ and $\epsilon(t)$ are the time traces of the laser phase and the field amplitude variation, respectively.

The phase noise is calculated by the beat signal in Figure~\ref{Overview}(b) following the standard procedure~\cite{Kester2009}. Using the measured single-ended phase noise $S_\phi(f)$ (in $\textrm{rad}^2/\textrm{Hz}$), we can calculate sets of the time trace $\phi^i(t)$ for the given laser phase noise:
\begin{equation}
    \phi^i(t)=\sum_f \sqrt{2 S_\phi(f) \Delta f} \textrm{cos}(2\pi f t+\phi^i_f)
,\end{equation}
where $\phi^i_f$ is the random phase offset of frequency $f$ and phase noise trace $i$, $\Delta f$ is the frequency resolution of $S_\phi(f)$. The phase offset is sampled uniformly, $\phi^i_f \in [0, 2\pi)$. 

By the same method, we can calculate the laser intensity noise trace $\Delta I^i(t)$ with the intensity noise spectrum $S_I(f)$ shown in Figure~\ref{Figure3}(d). Since the Rabi frequency $\Omega\sim\sqrt{I}$, $\epsilon(t)$ is determined by $\sqrt{[I_0+\Delta I^i(t)]/I_0}-1$, where $I_0$ is the average laser intensity. Considering the noise is wide-band and small ($\Delta I \ll I_0$), we can use the approximation $\sqrt{[I_0+\Delta I^i(t)]/I_0}-1=\frac{1}{2} \Delta I^i(t)/I_0$.

With the calculated time traces $\phi^i(t)$ and $\epsilon^i(t)$, we numerically simulate the time-dependent Hamiltonians using the Runge-Kutta method~\cite{Press2007}. We average the time evolution of the state vector over a series of $N=50$ trials of the phase noise trace. The Rabi dynamics and the corresponding fidelity under different Rabi frequencies are shown in Figures~\ref{Overview}(c) and~\ref{Figure4}. We note that the intensity noise and phase noise are uncorrelated in our simulation, which is not exactly true in the real setup. However, we believe this will only introduce very small difference since the intensity noise is significantly smaller than the phase noise.

\section{Details of the laser system}
\label{LaserSystem}

The laser we used in this experiment is an ECDL from Toptica Photonics working at 674 nm for Sr$^+$ ions. The laser is locked to a high-finesse ($\mathcal{F}\approx35\,000$) cavity provided by Stable Laser Systems using the Pound Drever Hall (PDH) technique and the cavity linewidth is around 45 kHz. (For a state-of-the-art high-finesse cavity with $\mathcal{F}\gtrsim100\,000$, the laser linewidth of the cavity is usually smaller than 10 kHz.) For the frequency stability of our system, the measured Allan deviation is $8\times10^{-15}$ for $\tau=1\textrm{ s}$, according to the documentation provided by Stable Laser Systems.  

Finally, we note that some PDH locking systems may involve the sideband locking technique, which gives a frequency difference between the laser and the cavity-transmitted light up to half of the cavity's free spectral range (FSR). Our method is also suitable for this tunable sideband technique by frequency down-converting (i.e., using an rf down converter or frequency divider) the beat signal to lower rf frequency. In fact, the ability to easily handle this offset is another advantage of our approach: if the cavity-transmitted light is used to interrogate atoms, the frequency offset must be addressed with modulators.

\bibliography{library}

\end{document}